\documentclass{PoS}

\usepackage{graphicx}

\title{Particle acceleration and radiation in Pulsar Wind Nebulae}

\ShortTitle{Particle acceleration in PWNe}

\author{\speaker{Elena Amato}\thanks{A footnote may follow.}\\
        INAF - Osservatorio Astrofisico di Arcetri, Largo E. Fermi, 5, I-50125, Firenze, Italy\\
        E-mail: \email{amato@arcetri.astro.it}}

\abstract{Pulsar Wind Nebulae are the astrophysical sources that host the most relativistic shocks in Nature and the only Galactic sources in which we have direct evidence of PeV particles. These facts make them very interesting from the point of view of particle acceleration physics, and their proximity and brightness make them a place where fundamental processes common to different classes of relativistic sources have a better chance to be understood.

I will discuss how well we understand the physics of Pulsar Wind Nebulae, describing recent progress and highlighting the main open questions. I will be mostly concerned with the subject of particle acceleration, but, as we will see, in order to clarify the physics of this process, it is important to determine the conditions of the plasma in the nebula. These in turn can only be constrained through detailed modelling of the PWN dynamics and radiation.

The shock in the Crab Nebula is probably the most efficient accelerator known, both in terms of conversion of the flow energy into accelerated particles (tens of percent) and in terms of maximum energy achieved ($\approx 10^{15}$ eV). I will review the different mechanisms proposed to explain particle acceleration and recent constraints derived from the comparison of synthetic emission maps with multi-wavelength data, including variability.}

\FullConference{Frontiers of Fundamental Physics 14 - FFP14,\\
                15-18 July 2014\\
                Aix Marseille University (AMU) Saint-Charles Campus, Marseille }
		
\begin{document}

\section{Introduction}
\label{sec:intro}
Pulsar Wind Nebulae (PWNe) are a class of Supernova Remnants (SNRs), whose broad-band emission is mostly non-thermal and powered by a fast-spinning magnetised neutron star, usually also observed as a pulsar. Most of the rotational energy lost by such a star goes into the acceleration of a highly relativistic magnetized wind, with a particle content dominated by electron-positron pairs. In the case of young pulsars, the wind is surrounded by the debris of the supernova explosion, which are in expansion at a much lower (non-relativistic) velocity. When the wind first impacts on the SNR, a reverse shock is launched towards the pulsar. At this shock (the termination shock, TS hereafter) the wind is slowed down and its bulk energy is efficiently converted into an outflow of relativistic particles that are then responsible for the nebular emission. Before discussing in more detail the current status of our understanding of such systems, let me briefly review the main reasons why these sources are of interest for several different fields of physics and astrophysics.

Pulsars are thought to be the primary leptonic antimatter factories in the Galaxy, but the exact amount of pair-production in their magnetospheres, the so-called pair multiplicity, $\kappa$, is not well established. PWNe shine with luminosities that are a substantial fraction of the pulsar total spin-down power, $\dot E$, much larger than what comes out in the form of pulsed emission. Therefore, modelling of these nebulae provides the tightest constraints on pulsar multiplicities \cite{arons12,nicco11} in a time when assessing the role of pulsars as positron producers is particularly topical, after PAMELA \cite{pamela} and AMS02 \cite{ams02} measured an anomalous rise of the $e^+/e^-$ ratio that does not find obvious explanation if the $e^+$ only come from cosmic ray interactions in the Galaxy. Possible interpretations of this anomaly involve dark-matter related processes, but their credibility can only be assessed after subtraction of all relevant astrophysical backgrounds, of which pulsars are likely the main contributors.

Understanding PWNe is important from the point of view of cosmic ray physics also in other respects. In these sources we observe the workings of the most relativistic shocks in Nature, an extreme version of the shocks that are usually invoked as particle accelerators to the highest energies observed, $\approx 10^{20}$ eV. PeV energies and acceleration efficiencies of tens of percent are inferred in the class prototype, the Crab Nebula, and they are reached in an environment that is highly hostile to particle acceleration according to standard theory. In fact, the most commonly invoked acceleration process in Astrophysics, the $1^{\rm st}$ order Fermi mechanism, or diffusive shock acceleration (DSA), does not work at highly relativistic shocks, unless the magnetic field is very weak \cite{sironi09}. A few alternative acceleration mechanisms have been proposed, but the viability of any of them depends on the flow magnetisation and composition. In order to understand particle acceleration it becomes then essential to unveil these two properties of the pulsar wind, which can only be assessed through detailed dynamical and radiation modelling of these systems. In the following I will discuss the current status of this kind of studies and their implications for particle acceleration mechanisms.

\section{The pulsar wind magnetisation}
\label{sec:sigma}
The best available model of PWN emission and dynamics is based on a magnetohydrodynamical (MHD) description of the pulsar outflow. The general idea is that the fast rotation of a magnetised neutron star induces an electric field that is strong enough to extract charges from the star. Electrons are certainly extracted. Whether also positive charges (protons or more generally ions) are extracted is an open question. Charge extraction is anyway insufficient to guarantee complete screening of the electric field parallel to the magnetic field everywhere in the star surroundings, so that the extracted electrons are accelerated in gaps of unscreened field and become energetic enough to produce pairs following the interaction with ambient photons, either thermal radiation from the star or self-produced while moving along curved magnetic field lines. The pair production process is not fully understood and well described, but the general consensus is that each electron extracted from the star generates a number $\kappa$ of pairs, with $10^3<\kappa<10^7$ in the case of young energetic pulsars \cite{arons12}. These pairs become then part of the cold relativistic outflow that is referred to as pulsar wind, and end up in the PWN. Indeed a direct estimate of the pulsar multiplicity $\kappa$ can be obtained from radiation modelling of PWNe: as we will see this exercise is not free of ambiguities, but estimates generally agree with the theoretical limits above. One important thing to notice is that if also ions are extracted from the star, these will end up to be part of the wind too. In addition, if $\kappa <m_i/m_e$, with $m_i$ and $m_e$ the ions' and electrons' mass respectively, the hadronic component of the wind carries more energy than the leptonic one. 

The pulsar wind is highly magnetized at its origin, but is thought to become matter dominated before approaching the termination shock: the ratio between Poynting flux and particle kinetic energy, $\sigma$, changes from $\sigma \gg 1$ to $\sigma\lesssim 1$. Again the process is not well understood. At large distances from the pulsar, the wind is estimated to be cold and have a bulk Lorentz factor $\Gamma$ in the range $10^4 -10^6$. Its structure, as inferred from analytical modelling and numerical simulations \cite{spit06}, is thought to be well approximated as a split monopole outflow: the wind expansion is radial, the magnetic field is predominantly azimuthal and decreasing with distance as $1/r$, and the energy flux is anisotropic, maximum in the equatorial plane of pulsar rotation and varying as $\sin^2\theta$, with $\theta$ the colatitude. In a region of angular extent equal to the inclination angle between the pulsar rotation and magnetic axes the wind is striped, meaning that regions of different field polarity alternate, separated by a current sheet. This is a place where magnetic reconnection is likely to take place \cite{klprev}.

Since the shock position is determined by the balance between the wind ram pressure and the pressure of the downstream plasma, which is relatively uniform, the anisotropy of the wind energy flux causes the termination shock surface to be oblate, closer to the star along the polar axis and at larger distance in the equatorial plane. This anisotropic energy outflow is at the origin \cite{lyub02} of the {\it jet-torus} morphology observed in the X-ray emission of the Crab Nebula and in a number of other nebulae. The interaction of the anisotropic pulsar wind with the surrounding SNR has been extensively studied by means of numerical simulations within the framework of 2D \cite{kl04,ldz04} (and more recently also 3D \cite{porth14}) relativistic MHD. These studies have provided important insights on the wind magnetisation at the shock front, which will be briefly summarised here.

Most of the numerical work has focused on the Crab Nebula, which is one of the best studied objects in the Universe, with data available over 20 decades in frequency, and high resolution images in the radio, optical and X-rays. The general assumption of these numerical studies is that particle acceleration occurs at the TS: accelerated particles are then advected with the MHD flow from their injection site into the nebula and meanwhile are subject to adiabatic and radiative losses. The latter are mainly due to synchrotron emission, but the Inverse Compton Scattering (ICS) process also needs to be taken into account in terms of radiation. The synchrotron emission spectrum is cut-off at a photon energy $\epsilon_{\rm sync}\approx 100$ MeV, and higher energy photons, detected from Crab up to $\epsilon_\gamma\approx 10$ TeV, are due to ICS: the same particles responsible for lower energy emission, upscatter the synchrotron emitted photons and external radiation fields (CMB and Infrared background) to higher energies. The different dependence on magnetic field of the two contributions allows to put tight constraints on the nebular magnetic field strength, by comparison of models with data. 

As far as the magnetisation at the shock is concerned, a lower limit can be set based on the the very existence of a X-ray jet. Within the MHD framework, the polar jet, showing flow speeds of order 0.5 c, results from the hoop stress associated with the toroidal magnetic field. Even if $\sigma<1$ at the termination shock, the magnetic field builds up in the immediate post-shock and unless $\sigma$ was too small it reaches energy equipartition with the particles. When this happens the flow is diverted towards the axis and the jet is formed.

Indeed 2D simulations show that in the case of the Crab Nebula a polar jet is formed as soon as $\langle \sigma\rangle>0.01$ averaged in latitude \cite{ldz04}. As we will discuss later this value of $\sigma$ is such that DSA cannot be at work along most of the shock front. For $\sigma\approx 0.025$ the morphology of the simulated nebula is very close to the observed one. However for such a value of $\sigma$ the above mentioned comparison between computed synchrotron and ICS emission with the data shows a discrepancy \cite{ldz06,volpi08}: the ICS emission is overestimated, indicating that the magnetic field in the simulated nebula is lower than in reality. Attempts at solving this discrepancy by increasing $\sigma$ result into emission maps with polar jets much brighter than the equatorial features, at odds with observations from the point of view of morphology.

A possible solution to this puzzle has been recently suggested by the results of the first 3D MHD simulations performed for the Crab Nebula: the exceedingly strong jets are likely an artefact of axisymmetry. In 3D, the development of kink instabilities (that cannot develop in 2D) leads to the conversion of a large fraction of the magnetic energy in the nebula into poloidal field lines, much reducing the magnetic tension, as predicted by \cite{beg98}. As a result, for a given value of the magnetisation at the shock, the polar jet is much weaker in 3D than in 2D, so that one can hope to reproduce both the spectrum and the high energy morphology with $\sigma\approx 1$. The 3D simulations, however, have been run, so far, only for about 100 years \cite{porth14}. Given the fact that magnetic energy is shown to be effectively dissipated (due to numerical resistivity), it is not clear that the average field strength will be as large as inferred from observations, $B\approx 150 \mu G$, when the system age is as large as 1000 yr, the Crab Nebula's actual age. Further investigation is clearly needed, but one conclusion that seems safe to draw from these preliminary results is that the actual value of the magnetisation at the shock can only be larger than what inferred from 2D simulations, $\langle\sigma\rangle>0.03$. This conclusion has important implications for particle acceleration.

\section{What is the process of particle acceleration?}
\label{sec:acc}
I already mentioned that the pulsar wind TS is in theory a hostile environment for particle acceleration and yet it accelerates particles with very high efficiency and up to very high energies. The resulting particle spectrum is a broken power-law, $N(E)\propto E^{-\gamma_e}$, with $\gamma_e\approx1.5$ at low energies and $\gamma_e\approx 2.2$ at high energies. The slope of the high energy spectrum is what one expects from DSA at a relativistic shock, but this process only works if the magnetisation at the shock is sufficiently low \cite{sironi09}, $\sigma\lesssim10^{-3}$, whereas we just saw that global MHD simulations of the Crab Nebula lead to infer $\langle \sigma \rangle>0.03$. According to current studies, only few \% of the wind energy flows though sectors of the TS where $\sigma \lesssim 10^{-3}$ \cite{hepro4}, which makes DSA unlikely as the main acceleration process.

Alternative proposals that have received some attention are: 1) driven magnetic reconnection at the TS; 2) resonant absorption of ion-cyclotron waves in a ion-doped plasma. The viability of both processes depends on the wind composition and pulsar multiplicity. Driven reconnection has recently been investigated \cite{sironi11} in a set- up appropriate to describe the Crab Nebula. The idea is that if the pulsar wind keeps its stripes with opposite direction of the magnetic field all the way to the TS, there compression causes the field to reconnect. A number of reconnection islands develop in the flow, where unscreened electric fields can effectively accelerate particles: the slope and the extension in energy of the resulting particle spectrum depend on the flow magnetization and on the ratio between the wavelength of the stripes and the particle Larmor radius. The latter ratio can be expressed in terms of $\kappa$ and the final result is that in order to reproduce the spectrum of radio particles, one would need $\sigma>30$ and $\kappa > 10^7$. Even ignoring the fact that such a high value of $\kappa$ is difficult to explain from the theory point of view, a wind with such a large number of pairs is actually likely to reconnect before reaching the TS (see \cite{hepro4} and references therein), causing $\sigma$ to decrease below the required minimum. The requirements on $\kappa$ would be less severe if the process took place at high latitudes, since here the shock is closer to the pulsar and the particle density scales as $r^{-2}$. However at high latitudes one does not expect any stripes.

An alternative proposal, that works for whatever $\sigma$, but requires that most of the energy of the pulsar wind be carried by ions, is that of resonant absorption by the pairs of the cyclotron radiation emitted by such ions. The idea is that at the crossing of the TS, the sudden enhancement of the magnetic field sets the plasma into gyration. The pairs quickly thermalize through emission and absorption of cyclotron waves, but ions with the same initial Lorentz factor (the wind is cold, so that all particles were moving with the same bulk Lorentz factor) react on time-scales that are longer by a factor $m_i/m_e$. If the wind is sufficiently cold ($\delta u/u < m_e/m_i$, with $u$ the flow four velocity)  before the TS, the ions emit waves with large power not only at the fundamental frequency of their gyration, but up to a frequency $m_i/m_e$ times higher, which can then be resonantly absorbed by the pairs. The resulting acceleration efficiency $\epsilon_{\rm acc}$, spectral slope $\gamma_e$ and maximum energy $E_{\rm max}$, all depend on the fraction of energy carried by the ions $U_i/U_{\rm tot}$. PIC simulations show a wide variety of values: $\epsilon_{\rm acc}=few (30)$ \%, $\gamma_e > 3 (<2)$, $E_{\rm max}/(m_i \Gamma c^2)=0.2 (0.8)$ for $U_i/U_{\rm tot}=0.6 (0.8)$ \cite{amato06}. Once again the pulsar multiplicity plays a crucial role, since ions can only come from the star, they are at most as numerous as primary electrons. Therefore, if $\kappa>m_i/m_e$, $n_i m_i \Gamma c^2<n_\pm m_e \Gamma c^2$ and ions cannot dominate the wind energy budget.  It is interesting to notice that if ions are there, instead, these would be PeV ions \cite{hepro4}.

\section{The pulsar wind composition}
\label{sec:compo}
The most direct way of estimating the pulsar multiplicity is from modelling of the nebular emission: when this is done, one finds that in the Crab Nebula radio emission requires $\kappa \approx 10^6$, while X-ray emission implies $\kappa \approx 10^4$. While X-ray emitting particles have very short synchrotron life-times and must be tracing the current pair injection rate by the pulsar, radio emitting particles suffer little losses and might in principle be fossil \cite{atah96}. Clarifying their origin is fundamental to obtain a correct estimate of $\kappa$ and constrain the process of particle acceleration.

A recent study has taken into account three different scenarios for the injection of radio emitting particles and tried to discriminate among them based on the emission morphology \cite{olmi14}. The result is that a scenario in which these particles are injected only for a short time after the SN explosion, with no further reacceleration, can actually be excluded. On the other hand, it does not seem possible to exclude any of the two other scenarios: A) ongoing injection at the TS followed by advection with the flow; B) uniform distribution in the nebula (as due to continuous reacceleration by spatially distributed turbulence in the body of the nebula). When synchrotron emission is computed on top of the MHD simulations, these two different hypotheses on the spatial distribution of radio particles produce very similar results: the emission maps are so similar that comparison with observations does not allow to discriminate between the two cases. 

In addition, also time-variability in the inner region is found in both cases and with very similar features. This is an important result, because the existence of variable radio features in the inner nebula, {\it radio wisps}, could be naively taken as evidence supporting ongoing injection of radio particles at the TS. In fact, wisp variability arises in simulated maps from the properties of the MHD flow: the moving bright features, seen at radio, optical and X-ray frequencies, are associated, in simulated maps, with locally enhanced magnetic field and Doppler boosting.

However there is more to this. Multi wavelength observation campaigns of the Crab Nebula have shown that the {\it wisps} are not coincident at the different wavelengths \cite{schwei13}. Within the MHD description of the flow such an effect can only result from differences in the acceleration site of particles of different energies. If synchrotron emission maps are computed at the various frequencies, after assuming different acceleration sites for particles of different energies, it is possible to gain some insight from multi wavelength variability. A recent effort in this direction \cite{olmi15} shows that if the MHD description holds, observations can only be explained if radio emitting particles are uniform in the Nebula or accelerated in a wide sector of the shock front, while higher energies are reached only within a smaller angular sector around the equatorial plane. The general picture that emerges is that the conditions for acceleration of radio particles must be met by most of the shock surface, while X-ray emitting particles might require more stringent conditions that are only reached around the equator. It is possible, for example, that indeed they can only be accelerated where the magnetisation is low enough. But this would have to occur for a larger fraction of the flow than currently thought.

\section{Summary and Conclusions}
\label{sec:concl}
Unveiling the mechanism of particle acceleration in PWNe is an especially challenging and fascinating problem, with implications for our general understanding of how the highest energies are reached in Nature. The answer to this question requires knowledge of plasma parameters that must be inferred from detailed modelling of the spectral and morphological properties of the emission, which so far has only been done for the Crab Nebula.

Current 2D MHD models, very successful at explaining the high energy morphology of the Crab Nebula, require too low $\sigma$ and lead to a nebular magnetic field which is too low to account for the combined synchrotron and ICS spectrum. The first 3D studies show that kink instabilities might play an important role in tangling the field, reducing the hoop stress and allowing to reproduce the morphology with a larger $\sigma$. Current 3D simulations however have too short a duration and probably also excessive dissipation. Further investigation is needed but a conclusion that seems unavoidable is that along most of the TS, $\sigma$ is too high for DSA to operate. 

The two alternative mechanisms that have been proposed have opposite requirements on the pulsar multiplicity: driven magnetic reconnection requires $\kappa$ to be larger than currently believed, while ion-cyclotron absorption can only work if ions are extracted from the star and $\kappa<m_i/m_e$. The main unknown in determining the value of $\kappa$ is the origin of the low energy (radio emitting) particles, which has proven so far very difficult to assess based on emission morphology and time variability: the only scenario that could be excluded is that of fossil particles with no reacceleration. More work is needed, and detailed studies of objects other than the Crab Nebula.

\acknowledgments{Attendance of this conference has been made possible by the grant PRIN-INAF 2012}

\end{document}